\documentclass[prb,showpacs,twocolumn]{revtex4}
\usepackage{graphicx}
\usepackage{subfigure}
\begin{document}

\title{Unconventional superconducting states on doped Shastry-Sutherland lattice}

\author{Huai-Xiang Huang$^1$, Yan Chen$^{2,3}$, Guo-Hong Yang$^1$}
\affiliation {$^1$ Department of Physics, Shanghai University,
Shanghai, 200444, China\\
$^2$ Department of Physics and Lab of Advanced Materials, Fudan
University, Shanghai, 200433, China\\
$^3$ Department of Physics and Texas Center for Superconductivity, University
of Houston, Texas 77204, U.S.A}

\date{\today}

\begin{abstract}
By using a renormalized mean-field theory, we investigate the phase
diagram of $t$-$t'$-$J$-$J'$ model on  two dimensional
Shastry-Sutherland lattice which are topologically equivalent to
synthesized material $\mathrm{SrCu_2(BO_3)_2}$. We find that the
symmetry of superconductivity ground state depends on the 
frustration amplitude $\eta=t'/t$ and doping concentration.
For weak to intermediate frustration, $d_{x^2-y^2}$-wave pairing symmetry is robust 
in a large parameter region. 
Around the symmetric point $|\eta|=1$, $d$-wave,
$s$-$s$-wave pairing as well as staggered flux may serve as ground state
by varying the doping level. There is a first-order transition between these distinct ground states.
For larger frustration $|\eta|>1$, the ground state has an $s$-$s$-wave symmetry for 
both hole and electron doping.

\end{abstract}


\maketitle

\section{Introduction}

\begin{figure}
\includegraphics[width=7cm]{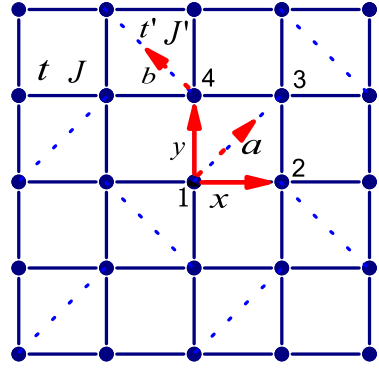}
\caption{Schematic structure of Shastry-Sutherland lattice. It includes
four sublattices  $1..4$. The hopping
integral and spin-spin coupling are $t$ and $J$ on the n.n.
links (solid line) and $t'$ and  $J'$ on  diagonal links (dashed
lines). We use $a,b$ to distinguish two diagonal links with different orientations.}\label{1}
\end{figure}

Geometrically frustrated lattices have crucial impacts on the emergence of exotic electronic states
in strongly correlated systems~\cite{poiblanc1,Aoki}, examples are triangular layered
cobaltates $\mathrm{Na_xCoO_2}$~\cite{co}, anisotropic triangular 
lattice $\mathrm{Cs_2CuCl_4}$~\cite{Cs}   and three dimensional
pyrochlore material $\mathrm{KOs_2O_6}$~\cite{kos}
In particular, resonating valance bond (RVB) spin liquid or valence
bond crystal may exist in frustrated quantum magnets. There is a hope that 
unconventional superconducting state may emerge upon doping of 
the frustrated magnets, it has been pointed out in the recent 
theoretical studies~\cite{YZhou02,Ogata03,Cs-Chung,did,Gan06,huang}.
Recent discovered  two dimensional synthesized frustrated
material~\cite{ss3} $\mathrm{SrCu_2(BO_3)_2}$ is an important
compound. It is topologically equivalent to the
Shastry-Sutherland~\cite{ss2,ss3} lattice,
spin-$\frac{1}{2}$ $\mathrm{Cu^{2+}}$ lies in two-dimensional
$\mathrm{CuBO_3}$ layers decoupled from each other by plane of
$\mathrm{Sr^{2+}}$ ions, the antiferromagnetic exchange couplings between
$\mathrm{Cu^{2+}}$ ions is identical to Heisenberg-hamiltonian of
$\mathrm{SS}$ lattice and motivate us to investigate its doping
properties. This lattice has been studied many years ago as a two
dimensional exactly solvable~\cite{ss1} spin model, a schematic
Shastry-Sutherland lattice is illustrated in Fig.~\ref{1}. 
Let $J$ and $J'$ be the exchange couplings along the square lattice and diagonal links, respectively. 
The production of valence-bond singlets  on disjointed diagonal links
is the exact ground state for $J'/J > 1.477$~\cite{ss8,ss81,ss82}, 
Experiments showed that $J'/J=1.574$ is an optimal
value~\cite{jpbj} for the insulator $\mathrm{SrCu_2(BO_3)_2}$.

There are many previous investigations on the doping effect of the Shastry-Sutherland 
lattice and various techniques have been used~\cite{ss5,ss4,ss6,ssh}.
By using slave-boson mean-field theory~\cite{ss5}, the competing orders of staggered flux state and 
d-wave superconducting state are investigated at a specific parameter regime. Similar results have been obtained
in a recent variational Monte Carlo study.~\cite{ss6}
Based upon the analysis of $t$-$J$-$V$ model via the bond-operator formulation, 
a number of superconducting states including $s$-wave, $(s+id)$-wave, plaquette 
$d$-wave are found as the ground states of the doped Shastry-Sutherland lattice.~\cite{ssh}
On the other hand, exact diagonalization approaches~\cite{ss4} have been employed to study the 
ground state of finite system and no superconducting order is found to be favored on doping.

In this paper, we apply the plain vanilla version of RVB
theory~\cite{Zhang88,vannila} to study the emergence of unconventional superconductivity. 
We define $\eta=t'/t$ as the frustration amplitude,
where $t'$ and $t$ are hopping integrals on diagonal links and  
square lattice links, respectively, and use $t$-$t'$-$J$-$J'$ model to
study the doping effect on the Shastry-Sutherland lattice. 
The competition among various superconducting states will be examined for 
both hole-doping and electron-doping cases.
The phase diagram is depicted as functions of $\eta$
and doping concentration $\delta$. In particular, four distinct
ground states show up. We classify these states
in terms of relative phase of mean-field pairing amplitudes.

There are four possible ground state candidates. 
In certain limiting cases such as $|\eta| \ll 1$, it is well known that pairing symmetry belongs
to $d_{x^2-y^2}$-wave with the superconducting order parameters on square lattice links $\Delta_x=-\Delta_y$ and 
the pairing parameters on diagonal links $\Delta_a=\Delta_b=0$.  Another candidate is the 
$s$-$s$-wave pairing symmetry with $\Delta_x=\Delta_y$ and 
$\Delta_a=\Delta_b$, while the relative phase shift between these two distinct links is equal to 
$\pi$. The third candidate state is 
staggered flux state, it can only be stable in negative $\eta$ and
small doping. In such state, the complex particle-hole mean-field parameter is 
modulated alternatively by a staggered magnetic $\pm\phi$. 
The last candidate is normal metal with vanishing of mean-field parameters.
Our calculation shows that from weak to intermediate frustration, 
$d$-wave state maintains in a large region of electron and hole dopings.
Around the symmetric point $|\eta|=1$, the symmetry of ground state 
is sensitive to the doping level since the energies of three distinct states, 
$d$-wave, $s$-$s$-wave pairing and staggered flux, are almost identical. 
For larger frustration $|\eta|>1$, the ground state has an $s$-$s$-wave symmetry for 
both hole and electron doping.

The rest of the paper is organized as follows. In Sec. II, we
propose the formalism of renormalized mean-field theory to study the 
$t$-$t'$-$J$-$J'$ model Hamiltonian on the Shastry-Sutherland lattice.
In Sec. III, we present our numerical results of renormalized mean-field theory as functions of 
frustration and doping level, and mean-field phase diagram as well. 
Finally a summary is given in Sec. IV.

\section{Formalism}

A primitive unit cell of the Shastry-Sutherland lattice includes four
inequivalent sites, we consider a  $t$-$t'$-$J$-$J'$ model on such lattice. 
The Hamiltonian can be written as
\begin{eqnarray}
H = &-&\sum_{\langle ij \rangle \sigma} t_{ij} \hat{P}(c_{i\sigma
}^{\dagger }c_{j\sigma }+ h.c.)\hat{P}+ \sum_{\langle ij
\rangle}J_{ij}
\vec{S}_{i}\cdot \vec{S}_{j} \nonumber \\
 & - & \mu\sum_{i} n_{i},
\end{eqnarray}
where $c_{i\sigma }^{\dagger }$ is to create a hole with spin
$\sigma $ at site $i$, $\vec{S}_{i}$ is a spin operator, $\mu$ is
the chemical potential, $\langle ij \rangle$ denotes a square lattice or diagonal link
on the lattice, $t_{ij}$ and $J_{ij}$ stand for the
hopping integrals and antiferromagnetic exchange couplings,
respectively, $t_{ij}=t$ and $J_{ij}=J$ on the square lattice links, while
$t_{ij}=t'$ and $J_{ij}=J'$ on the diagonal links, as shown in
Fig.~1. We use $t$ as an energy unit and set $t/J=3$ . We choose 
$J'/J=(t'/t)^{2}$ to be consistent
with the superexchange relation of $J= 4t^2/U$ in the large Hubbard
$U$ limit. Projection operator~\cite{Zhang88,vannila}
$\hat{P}=\prod\limits_{i}(1- n_{i\uparrow }n_{i\downarrow })$
removes all the doubly occupied states.

We define particle-particle condensate mean-field as well as particle-hole condensate
mean-field as,
\begin{eqnarray}
{\Delta}_{ij} &=& \langle
c^{\dag}_{i\uparrow}c^{\dag}_{j\downarrow}-
c^{\dag}_{i\downarrow}c^{\dag}_{j\uparrow}\rangle_0 \nonumber
\\
\xi_{ij} &=& \langle c^{\dag}_{i\uparrow}c_{j\uparrow}+
c^{\dag}_{i\downarrow}c_{j\downarrow}\rangle_{0},
\end{eqnarray}
where $\langle \rangle_0$ gives expectation value corresponding to
states without constraint of no double occupancy. Although the
number of independent parameters in Shastry-Sutherland lattice is twelve, our
calculation shows the number can be reduced to eight due to certain symmetry.
The effect of the projection operator is taken into account by a set
of renormalized factors~\cite{gutzwiller,vallhardt}, which are
determined by statistical countings. Within the Gutzwiller
approximation, the energy of physical state $|\psi\rangle$ can be
reduced to that of state $|\psi_0\rangle$ which is free of double
occupancy constraint, i.e.,
$\langle\psi|H|\psi\rangle=\langle\psi_0|H'|\psi_0\rangle=\langle\psi_0|g_tH_t+g_sH_s|\psi_0\rangle$.
In homogenous case the renormalized factors  $g_{t}=2\delta
/(1+\delta)$ and $g_{s}=4/(1+\delta)^2$,  where $\delta$ denotes
the doping density. Thus, we have the effective Hamiltonian,
\begin{eqnarray}
H_{eff} & = & \sum_{\langle ij \rangle \sigma} -g_t t_{ij}
(c_{i\sigma }^{\dagger }c_{j\sigma } + h.c) +\sum_{\langle ij
\rangle} g_s J_{ij}
\vec{S}_{i}\cdot \vec{S}_{j} \nonumber \\
& - & \mu\sum_{i} n_{i},
\end{eqnarray}
and the resulting mean-field Hamiltonian can be expressed as 
\begin{eqnarray}\label{effh}
H_{MF}&=&\sum_{ \langle ij \rangle \sigma}-\frac{3}{8} g_s J_{ij}
[\xi _{ij} c_{i\sigma }^{\dagger }c_{j\sigma } + \Delta _{ij} c_{i
\sigma}^{\dagger }c_{j \bar{\sigma}} + h.c. ]\nonumber\\
&\,&-g_t t_{ij} (c_{i\sigma }^\dag  c_{j\sigma } + h.c) + const,
\end{eqnarray}
with $const=\frac{3}{8}Jg_s \sum_{\langle ij\rangle}[{ |\xi _{ij}|
}^2 + | \Delta _{ij} |^2]$. We diagonalize the mean-field
Hamiltonian (\ref{effh}) in momentum space, all the local
order parameters and the chemical potential $\mu$ are 
self-consistently obtained for each set of frustration parameter $\eta$  
and doping density $\delta$, with this procedures the lowest energy state
can be determined.

\section{Numerical Results of Phase Diagram and Mean-field Theory}

In this section, we present our numerical results of renormalized mean-field theory 
on the Shastry-Sutherland lattice. The mean-field order parameters depend on both 
frustration parameter $\eta$ and doping level $\delta$. 
In our calculations, we choose several typical frustration amplitude 
to analyze pairing symmetry for different doping
levels. Larger frustration parameter $|\eta|$ corresponds 
to stronger interactions on the diagonal bonds.
and the symmetric point $|\eta|=1$ has the strongest frustration.
We will start from the phase diagram, then provide detailed discussion of mean-field order 
parameters as functions of frustration parameter $\eta$ and doping level $\delta$.

%\subsection{Phase diagram}

%{\it Phase Diagram} 
\begin{figure}
\includegraphics[width=8cm]{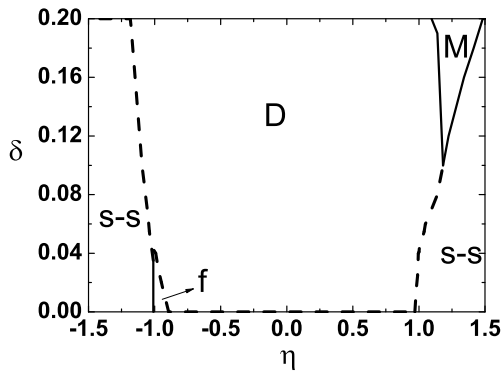}
\caption{Phase diagram of $t$-$t'$-$J$-$J'$ model on the Shastry-Sutherland 
lattice as functions of doping density $\delta$ and
frustration amplitude $\eta$. The thick solid line denotes a first order
phase boundary while the dashed line corresponds to a second order
transition.}\label{2}
\end{figure}

As shown in Fig. \ref{2}. there exists four distinct phases in the phase diagram. 
It is obvious that the ground state has a $d_{x^2-y^2}$-wave or $d$-wave symmetry in the limit of 
$\eta \ll 1$ at finite doping. Our results show that $d_{x^2-y^2}$-wave state is stable in a wide 
parameter regions of $-\sqrt{0.9}<\eta<\sqrt{0.96}$ and finite doping. 
Previous studies have shown the robustness of $d$-wave pairing against 
weak frustrations on both triangular lattices and checkerboard lattices.~\cite{Ogata03,did,Gan06,huang}.
It seems that such robustness is universal for weakly frustrated systems. 
At large $|\eta|>1$, the ground state has an $s$-$s$-wave pairing symmetry 
with $\Delta_x=\Delta_y$ and $\Delta_a=\Delta_b$ while the relative phase between $\Delta_x$
and $\Delta_a$ is $\pi$. 
Recently the two families of the Fe-based superconductors are 1111 systems ReOFeAs with rare earth ions 
Re~\cite{Kamihara} and the 122 systems AeFe$_2$As$_2$ with alkaline 
earth element Ae~\cite{Rotter}. An $s$-$s$-wave pairing symmetry was proposed as a popular 
candidate for the superconducting pairing symmetry of the Fe-based superconductors.~\cite{Mazin}.

In between the above two regions, there are two non-superconducting states 
in such small parameter region around $|\eta|=1$.
The region around $\eta=-1$ corresponds to staggered flux state at low doping 
while the normal metal state prevails for $\eta \geq 1.2$ at finite doping ($\delta > 0.10$).
It is interesting to find that there is an abrupt change of superconducting order parameters in between 
$d$-wave and $s$-$s$-wave state around $\eta=-1$ and the phase transition is first order. 
Around $\eta=1$ region,  phase transition from $s$-$s$-wave to $d$-wave state is a 
weakly first-order transition in which parameters change continuous at the boundary.
Moreover, the phase transition between staggered-flux
state and $d$-wave state is also  first-order. Other phase boundary corresponds to second-order.

%\subsection{first-order phase transition for $|\eta|\leq1$ }

\begin{figure}
\includegraphics[width=9cm]{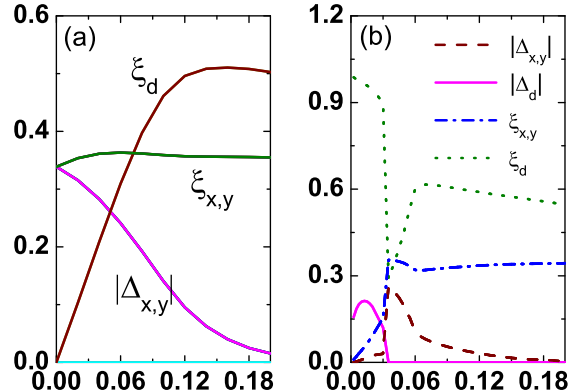}
\caption{The magnitudes of the mean-field order 
parameters $\Delta$ and $\xi$ as functions of  
$\delta$ for (a) $\eta=\sqrt{0.8}$ and (b) $\eta=1$. }\label{3}
\end{figure}

As we pointed out already, in the limit of weak frustration $|\eta| \ll 1$ upon doping,
the model Hamiltonian may correspond to
the well-known $t$-$J$ model in which the $d$-wave superconducting symmetry is the ground state.
Our calculations are performed for various frustration parameter as well as doping level.
In a wide range of parameter region, $d$-wave state appears to be robust as the ground state. 
In particular, our calculations show that $d$-wave state have lowest energy 
for positive $\eta$ less than $\sqrt{0.96}$. 
In Fig. 3, we present the amplitudes of the mean-field parameters as functions of hole density $\delta$ 
for $\eta=\sqrt{0.8}$ and $\eta=1$, respectively.
As shown in Fig. 3(a) for $\eta=\sqrt{0.8}$, a typical $d$-wave state is obtained and the parameter $\xi_d$ shows
no much doping dependence.
In the parameter region $0.96<\eta\leq1$, $d$-wave and $s$-$s$-wave superconducting 
state are highly competing. The mean-field order parameters of ground state are discontinuous
as functions of hole density $\delta$. For better illustration, we take the symmetric point $\eta=1$. 
As displayed in Fig. 3(b), the ground state has $s$-$s$-wave symmetry at small doping while the
$d$-wave state prevails for larger doping level. The critical doping level corresponds to $\delta_c \simeq 0.035$.

\begin{figure}
\includegraphics[width=9.5cm]{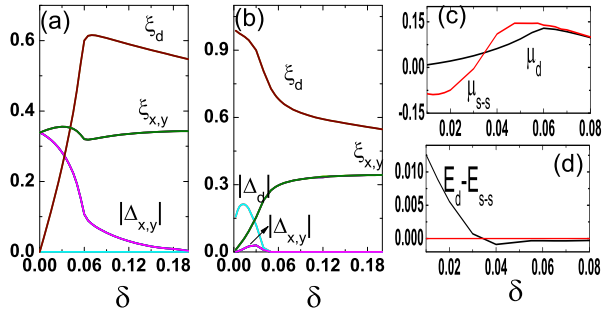}
\caption{Panels (a) and (b) describe the mean-field order parameters as functions
of $\delta$ for $d$-wave and $s$-$s$-wave state for $\eta=1$, respectively.
Panels (c) and (d) correspond to the evolutions of chemical potential and energy per site as a
function of  $\delta$ for two competing states. }\label{4}
\end{figure}

To reveal the competition between $s$-$s$-wave and $d$-wave
states more clearly, we compare the mean-field order parameters, 
chemical potential  as well as energy per site for these two states
in Fig. 4 at the symmetric point  $\eta=1$. 
Fig. \ref{4}(a) shows parameter functions of $d$-wave,
Fig.\ref{4}(b) shows that for $s$-$s$-wave state in which
$|\Delta_d|$ is larger than $|\Delta_{x,y}|$ where the subscript
$d$ denote the diagonal bonds. For $s$-$s$-wave, all pairing
parameters change non-monotonically to zero, and then metallic
state emerges smoothly. We plot the parameters of $s$-$s$-wave
from $\delta=0.005$, at half filling there is no
self-consistent $s$-$s$-wave solution. Fig. \ref{4}(c) 
shows the crossing of chemical potentials for those two competing
states at the transition point $\delta\simeq0.035$.
In Fig. ref{4}(d), such a crossing of energy per site for
those two states exhibits itself as well.
It is rather clear that a zero-temperature first-order quantum
phase transition may occur at the transition point $\delta\simeq0.035$.

%In Fig. \ref{5}, for positive $\eta=1$, we plot the constant
%energy curves of tight band $\mathrm{SS}$ lattice. The
%four-sublattice structure leads to four energy bands, calculation
%shows that the first band is fully filled, the forth band is
%empty, the second band filled with percentage of $57.8\%$ and the
%third one filled with percentage of $42.2\%$ which are shown in
%Fig. \ref{5}(a),(b) respectively.
%\begin{figure}
%\includegraphics[width=9cm]{Fig5.eps}
%\caption{(a) shows the constant energy curve of the second energy
%band, and (b) shows the constant energy curve of the third energy
%band.}\label{5}
%\end{figure}

\begin{figure}
\includegraphics[width=9cm]{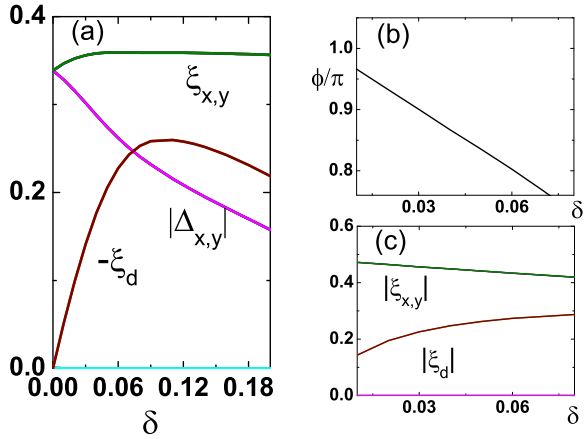}
\caption{Amplitudes of the mean-field order parameters as functions of 
$\delta$ for $\eta=-\sqrt{0.95}$. Panels (a) and (c) correspond to 
$d$-wave and staggered flux state, respectively. Panel (b) describes 
the accumulated phase of $\xi$ for staggered flux state.}\label{5}
\end{figure}

From Fig. \ref{2}, one can see that near $\eta=-1$, a staggered-flux state may appear
in a small parameter region. For instance,  we plot the mean-field parameters of $d$-wave state and
staggered-flux state for $\eta=-\sqrt{0.95}$ in Fig.
\ref{5}(a),(c) respectively. Calculation shows that for staggered
flux state $\xi_{ij}$ is a complex value. The phase of $\xi_d$ is
$\pi$, and the accumulating phase of $\xi_{ij}$  on a pane is
independent of $\eta$ and reduces linearly with increasing doping,
as shown in Fig. \ref{5}(b). At half-filling, it is hard to obtain
self-consistent solution.  For $\eta$ less than $0.04$, 
the staggered flux state is stable, while in high doping level $d$-wave
state has lower energy. For $\eta<-1$ $s$-$s$-wave state emerges
with introduce of mobile charge, and have favorable energy than the
staggered flux state.

%\subsection{weakly first-order phase transition for $\eta>1$}
\begin{figure}
\includegraphics[width=9cm]{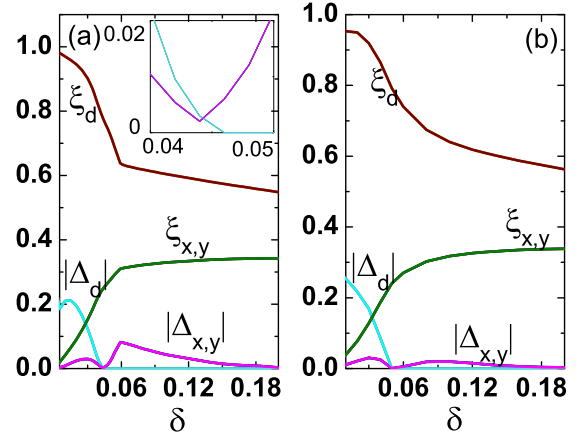}
\caption{Amplitudes of the mean-field order parameters as functions of 
$\delta$ for (a) $\eta=\sqrt{1.005}$ and (b) $\eta=\sqrt{1.08}$. 
The inset of (a) zooms in  the low doping region.}\label{6}
\end{figure}

For large frustrated amplitude, the interactions on diagonal bonds
may play a dominate role in determination of superconducting pairing symmetry. 
When $\eta$ takes a value slightly larger than $1$, as
the superconducting pairing symmetry may change from $s$-$s$-wave to $d$-wave 
and mean-field parameters varies rather smoothly. This transition is weakly first order.
In Fig. \ref{6}(a), it shows that $\Delta_d$ varies nonmonotonically to zero and
ground state evolves from $s$-$s$-wave state to $d$-wave state with
increasing doping for $\eta=\sqrt{1.005}$.
Precise calculation of pairing parameters shows that around the
critical point $\Delta_{x,y}\neq0$. This is illustrated in the
inset picture of Fig. \ref{6}(a), and indicates that this is a
weakly first order phase transition. Fig. \ref{6}(b) presents the 
mean-field parameter as functions of $\delta$ for
$\eta=\sqrt{1.08}$. We find that a larger $\eta$ corresponds to a
smaller amplitude of $\Delta_{x,y}$ of $d$-wave state. For
$\eta\geq1.4$, amplitude of the $d$-wave state is vanished and metal
state follows the $s$-$s$-wave.

%\subsection{cases for larger frustration amplitude}

\begin{figure}
\includegraphics[width=9cm]{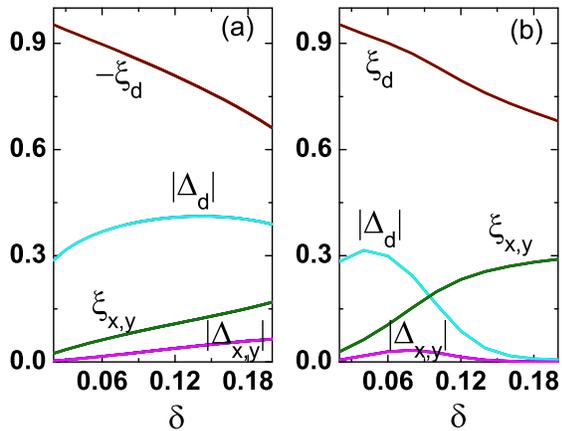}
\caption{Amplitudes of the mean-field order parameters as functions of 
$\delta$ for (a) $\eta=-\sqrt{2}$ and (b) $\eta=\sqrt{2}$.}\label{7}
\end{figure}

For large frustrated amplitude, $s$-$s$-wave state is the ground
state for both positive and negative $\eta$. Fig. \ref{7}(b) takes
$\eta=\sqrt{2}$ as an illustration. By increasing doping to a
considerable high level, both $|\Delta_{x,y}|$ and $|\Delta_d|$
approach zero. However they do not reach zero simultaneously and
$\Delta_{x,y}$ decreases more rapidly. It implies that
superconducting order parameter may exist only on diagonal bonds in some cases.
As positive $\eta$ becomes larger, no metal state will appear since the
pairing parameters of $s$-$s$-wave state may have finite amplitude at
high doping level.
For negative larger frustration amplitude cases, $s$-$s$-wave
state is ground state at all doping level. Amplitude of $\Delta_d$
is larger comparing to that of corresponding positive case, since
the negative $t'$ frustrated hopping thus enhance pairing
amplitude. It indicates that superconductivity favors electron
doping. This has been shown in Fig. \ref{7}(a). It should be pointed out
that for larger frustrated case, our mean field theory can not
obtain the exact dimer ground state at half-filling.
%Fugacity~\cite{fug1,fug2,fug3} of pair introduced by Laughlin can
%help understand the shortcoming of this mean field theory. 

\section{Summary}

We have employed the renormalized mean-field theory to study the
$t$-$t'$-$J$-$J'$ model on the geometrically frustrated Shastry-Sutherland 
lattice for both hole and electron doping cases. Our
calculation shows that the ground state of the doped system 
depends on the frustration amplitude $\eta$ and doping
level $\delta$.  For weak frustration $\eta << 1$, $d$-wave state is 
stable in a large parameter region in agreement with the case of $t$-$J$ model on
square lattice.  
For strong frustration $\eta > 1$, $s$-$s$-wave state
dominates in a wide range of parameter region. This feature 
has also been found in the doped triangular and checkerboard antiferromagnets. 
When approaching the most frustrated point
$\eta=1$, $d$-wave state competes with $s$-$s$-wave state, the
phase transitions are first-order, the parameters change suddenly
at critical point. Near $\eta=-1$, staggered flux state dominates.
For frustrated amplitude is not very large, as doping increasing
ground state changes from $s$-$s$-wave state to $d$-wave state via
the weakly first-order transition.
Moreover, we have  found the enhancement of superconducting order parameter for
negative $\eta$  because the negative $t'$ may introduce
frustration in kinetic energy and result in the enhancement of the pairing amplitude.
Our theoretical predications might be examined in future experiments on doped 
SrCu$_2$(BO$_3$)$_2$.

\section{Acknowledgments}

H.X.H. would like to thank Profs. F.C. Zhang, Y.Q. Li and Y. Jiang for
helpful discussions. This work was supported by the National Natural 
Science Foundation of China (Grants No. 10747145 and No. 10874032) and 
the State Key Programs of China (Grant No. 2009CB929204). 
Y.C. acknowledges the support from Shanghai Municipal Education Committee.

\end{document}